\documentclass[a4paper]{article}

\usepackage[latin1]{inputenc}
\usepackage[english]{babel}
\usepackage[final]{graphicx}
\usepackage{verbatim}

\usepackage{amsmath,amsfonts,amssymb,amsthm}
\usepackage{latexsym}

\newcommand{\structure}[1]{$^{#1}$}

\newcommand{\authorstructure}[2][]{$ $\hspace{-5mm}$^{#1}$ {\small \it #2}}

\newenvironment{Aabstract}
	{\vspace{5mm}
	 \begin{center}\bf Abstract\end{center}
	 \begin{quote}\small}
	{\end{quote}\vspace{5mm}}

\begin{document}

\begin{center}
{\LARGE Data Management and Mining \mbox{in Astrophysical Databases}}
\\[5mm]
{\sc Marco Frailis\structure{a,b},
Alessandro De Angelis\structure{a,b}, 
Vito Roberto\structure{c}}
\end{center}

\authorstructure[a]{Dipartimento di Fisica, 
                   Universit\`a di Udine, 
                   via delle Scienze~208, 33100~Udine, Italy}

\authorstructure[b]{INFN,\:Sez.\:di Trieste,\:% 
                   Gruppo Collegato di Udine,\:%
                   via delle Scienze\:208,\:33100~Udine, Italy}

\authorstructure[c]{Dipartimento\:di\:Matematica\:e\:Informatica,\:% 
                   Universit\`a\:di\:Udine,\:% 
                   via\:delle\:Scienze\:208,\ 33100\:Udine,\:Italy}

\begin{Aabstract}
  We analyse the issues involved in the management and mining of astrophysical
  data. The traditional approach to data management in the astrophysical field
  is not able to keep up with the increasing size of the data gathered by modern
  detectors. An essential role in the astrophysical research will be assumed by
  automatic tools for information extraction from large datasets, i.e. data
  mining techniques, such as clustering and classification algorithms. This asks
  for an approach to data management based on data warehousing, emphasizing the
  efficiency and simplicity of data access; efficiency is obtained using
  multidimensional access methods and simplicity is achieved by properly
  handling metadata.  Clustering and classification techniques, on large
  datasets, pose additional requirements: computational and memory scalability
  with respect to the data size, interpretability and objectivity of clustering
  or classification results. In this study we address some possible solutions.
\end{Aabstract}

%=============================================================
\section{Introduction}
%=============================================================

Data mining is the exploration and analysis, by automatic or semiautomatic
means, of large quantities of data in order to discover meaningful patterns and
rules~\cite{BL97}. Its goal is to find \emph{patterns} or \emph{relationships}
in data using techniques to synthesize models which are abstract representations
of reality. There are two main types of data mining models~\cite{Bor00}:
\emph{descriptive} and \emph{predictive}. \emph{Descriptive} models represent
patterns in data and are generally used to build meaningful groups or clusters.
\emph{Predictive} models are used to forecast explicit values, on the basis of
samples with known results.

At present, astrophysics is a discipline in which the exponential growth and
heterogeneity of data require the use of data mining techniques.  Due to the
advances in telescopes, detectors and computer technologies, astrophysics has
become a domain extremely rich of scientific data. The primary source of
astronomical data are the systematic sky surveys over a wide photon energy range
(from $10^{-7}$ eV to $10^{13}$ eV) \cite{BDPS01}. Large archives and digital
sky surveys with dimensions of $10^{12}$ bytes currently exist, while in the
near future they will reach sizes of the order of $10^{15}$ bytes. Numerical
simulations are also producing comparable volumes of information.

Therefore, the use of data mining techniques is necessary to maximize the
information extraction from such a growing quantity of data. Data
mining has reached a certain degree of maturity in data exploration for decision
support mostly in domains like marketing, sales and customer care. In the
astrophysics domain the role of data mining is to help researchers building or
verifying new physical models based on observational data.

%% This will enable \emph{qualitatively} and \emph{quantitatively} a new science,
%% ranging from statistical analysis of our galaxy and the large-scale structure of
%% the universe to the discovery of new or rare astronomical objects and phenomena
%%~\cite{BDPS01}.

A first issue in applying these techniques is the heterogeneity of astronomical
data, due in part to their high dimensionality including both spatial and
temporal components, due in part to the multiplicity of instruments and
projects. Another issue is that currently astronomical data are organized in a
traditional operational system, in which the emphasis is on data normalization;
data mining techniques, instead, require putting more emphasis on efficiency and
simplicity of data access, even if this entails some redundancy.  Gathering data
from multiple astronomical datasets to perform multi-wavelength analysis
necessitates both using an informational system, or data warehouse, as a model
for data management, and the definition of a common set of metadata to guarantee
the interoperability between different archives and a simpler data exploration.

Data mining techniques are rather general and can be employed in different
application domains in which there is an intensive use of data. Such techniques
include~\cite{BDPS01}:
\begin{enumerate}
\item Clustering techniques, such as Expectation Maximization (EM)
  with mixture models or Self-Organizing Maps (SOM), to find regions
  of interest, produce descriptive summaries and build density
  estimates of large datasets, or methods like Support Vector
  Clustering (SVC) to single-out rare objects or anomalous events.
\item Classification techniques, such as decision trees,
  nearest-neighbor classifiers, neural networks and statistical
  learning methods like Support Vector Machines (SVM), to categorize
  objects or clusters of objects of interest. The classification result
  is further analyzed to verify whether physically meaningful objects or
  groups of objects have been identified, and if these objects are
  present in some catalog or they are new.
\item Techniques to improve clustering and classification algorithms, such as
  genetic algorithms and Principal Component Analysis (PCA). These methods allow
  to find the best features and reduce the dimensionality of the domain
  space.
\item Software agents for automatic or semi-automatic information search and
  analysis.
\item Data visualization and presentation techniques (exploratory data
  mining). These techniques allow to present multidimensional information in a
  way easily understandable by a human user.
\end{enumerate}

%=============================================================
\section{Data management in astrophysical databases}
%=============================================================
At present, several multi-wavelength projects are underway, for example SDSS,
GALEX, \mbox{GSC-2}, POSS2, ROSAT, FIRST and DENIS. In the next years new
spatial missions will be launched; two of them, AGILE and GLAST, will observe
gamma-rays on a wide energy range. Besides, ground based telescopes sensitive to
gamma-rays are being tested or starting taking data (MAGIC, HESS, VERITAS,
CANGAROO III).

Therefore, astrophysicists will need a uniform interface to access all these
data~\cite{SKT00}.  Data gathered by all missions are heterogeneous as
they are mission oriented and dependent on the particular platform or instrument
(including hardware components information, quality flags decided inside the
mission, derived measures for particular analysis). Several scientific
research fields require to perform the analysis on multiple energy spectra and
consequently to get the data from different missions. Typically, astrophysicists
want to retrieve multi-spectral data for specific objects, classes of objects
(i.e. AGN, HII region) or selected regions of the sky. However, metadata in
mission archives are not designed to answer these queries~\cite{Che98}.

%=============================================================
\subsection{The data warehousing approach}
%=============================================================
Most of the online resources available to the astrophysicists community are
simple data archives. Typically, users can perform queries based on observational
parameters (detector, type of the observation, coordinates, astronomical object,
exposure time, etc.) to obtain images which are then processed by standard
analysis tools. Many astronomical catalogs can be accessed online, even if it is
still difficult to correlate objects in different archives or access multiple
catalogs simultaneously. Some advances, in this direction, have been
accomplished by projects like Vizier, Aladin and SkyView \cite{Viz,Alad,Sky}.

To identify objects and parameters which allow to answer directly to particular
scientific issues it is necessary to build a \emph{scientific archive} containing
the results of data analysis - scientific measurements - rather then the data
itself~\cite{DSZDG00}.  The users of this archive should be able to perform
queries based on scientific parameters (magnitude, redshift, spectral indexes,
morphological type of galaxies, etc.), easily discover the object types
contained into the archive and the available properties for each type, and define
the set of objects which they are interested in by constraining the values of
their scientific properties along with the desired level of detail.

Data mining applied to large astrophysical databases can involve the execution
of complex queries and multiple scans of large quantity of data. Therefore, it is
opportune to put more emphasis on data access efficiency rather than on data
normalization.

All the aforesaid requirements can be satisfied organizing data in a data
warehouse. A data warehouse can be defined as a \emph{subject-oriented},
\emph{integrated}, \emph{time varying} and \emph{non-volatile} data collection
\cite{Inm97}.  Subject-oriented means that in a data warehouse data are
collected and organized with the aim of a particular analysis. The second
property is surely the most important one; in fact, a data warehouse has to
integrate with the multiplicity of standards used by the sources it gathers data
from (i.e.  multiple astronomical catalogs). This data integration process can
involve conversion of types, formats or units and the addition of derived types
(i.e.  several statistical measures). A data warehouse is time varying because
its time horizon usually oscillates between 5 and 10 years and along this period
of time data collected are a series of snapshots taken at fixed times
\cite{Giu01}. It is non-volatile because data updating, and the resulting loss
of information, doesn't take place within it.

In a data warehouse, data are arranged in a structure that can be easily
explored and queried, with fewer tables and keys than the equivalent relational
model. You start from a relational model, but some restrictions are introduced by
using \emph{facts}, \emph{dimensions}, \emph{hierarchies} and \emph{measures} in
a characteristic star structure called \emph{star schema}~\cite{Pet94}. The
central table is called ``fact'' table and it is the highest dimensional table of
the scheme. It can represent a particular phenomenon that we want to study. This
table is surrounded by a number of tables, called ``dimensions'', which
represent entities related to the phenomenon to be studied and connected to the
central table, forming the ends of the star. Within the dimensions, attributes
are arranged in hierarchies, determining the ``drill-down'' and ``roll-up''
operations available on each dimension: the result is a tree that the user can
visit from the root to the leaves, refining his query (drill-down) or
generalizing it (roll-up).

%=============================================================
\subsection{The metadata role}
%=============================================================
Accessing data into a set of continuously evolving catalogs states the problem
of accessing and understanding parameters available in each catalog. A typical
problem is understanding whether a catalog contains some specific data type,
what is the reliability of these data, if they are written in a standard format,
if they are taken from other publications or catalogs, how the associated data
file can be processed. All these details describe data - they are metadata - and
traditionally are presented in the introduction of printed catalogs or detailed
in several publications analyzing the catalog data.

Metadata play an important role: a researcher has to obtain information about
the environment in which data have been gathered in order to understand the
respondence to the project requirements; for instance: date and/or data
acquisition method, internal or external error estimates, aim of data. Computing
systems have to access metadata to merge or compare data from different sources.
For instance, it is necessary that units are expressed unambiguously to allow
comparisons between data with different units.

The astrophysicists community, in addition to using the FITS (Flexible Image
Transport System) exchange format, is currently considering alternatives like
XML. Some attempts to define a common standard are XSIL (eXtensible Scientific
Interchange Language), XDF (eXtensible Data Format) and VOTable \cite{XSIL,XDF,VOTable}.

%=============================================================
\section{Spatial and multidimensional data structures}
%=============================================================
Spatial DataBase Systems (SDBS) are designed to handle spatial data and the
associated non-spatial information. Spatial data are characterized by a complex
structure (a spatial object can be a single point or a set of polygons
arbitrarily distributed). They are usually dynamic (requiring robust data
structures for frequent insertions, deletions and updates), tend to be large
(sky maps can reach sizes of Terabytes) requiring the integration of the
secondary storage. There is no standard spatial algebra, that is the set of
spatial operators depends on the specific application. Another important
property is that there is no total ordering on spatial objects preserving
spatial proximity~\cite{GG98}. This characteristic makes difficult to use
traditional indexing methods, like B-trees or linear hashing.

\emph{Spatial data mining} analyzes the relationships between the attributes of
a spatial object stored into the database and the attributes of the neighboring
ones. Typical queries required by this kind of analysis are: \emph{point
  queries}, to find all objects overlapping the query point; \emph{range
  queries}, to find all objects having at least one common point with a query
window; \emph{nearest neighbor queries}, to find all objects that have a
minimum distance from the query object.

These\,principles\,can\,be\,generalized\,to\,multidimensional\,data.\ Access\,\mbox{methods} to
multidimensional databases can be classified in \emph{Point Access Methods}~(PAM)
and \emph{Spatial Access Methods} (SAM). PAM are designed to perform searches on
point databases. They usually arrange data in buckets, each one corresponding to
a disk page. The buckets are indexed by either flat or hierarchical data
structures: flat structures are used in multidimensional hashing methods like
the grid file and EXCELL; hierarchical structures are used in hierarchical
access methods like quadtree, KD-tree and KD-B-tree. SAM manage objects with
spatial properties like area and shape. Access methods in SAM are often
extensions of PAM ones to handle objects with a spatial extent. Such methods include
R-tree, R$^*$-tree and Multi-layer grid file.

%=============================================================
\subsection{Quadtree}
%=============================================================
The term \emph{quadtree} is used to describe a class of hierarchical data
structures based on the principle of recursive decomposition of the space. They
can be distinguished by the following elements~\cite{Sam90}:
\begin{itemize}
\item the type of data they are used to represent;
\item the principle guiding the decomposition process;
\item the resolution (variable or not).
\end{itemize}
Until recent experiments, astronomical observations were restricted to a
selected region of the sky, making a planar projection of the observed
region adequate. However, the next generation experiments, like GLAST and AGILE, will provide
a detailed observation of the whole sky and thus they will require the handling of data with a
spherical distribution. In the astrophysical field, two methods for indexing the
sky based on quadtrees have been designed: the Hierarchical Triangular Mesh
(HTM) and the Hierarchical Equal Area isoLatitude Pixelization (HEALPix).

\begin{figure}
  \begin{center}
    \includegraphics[width=0.3\textwidth]{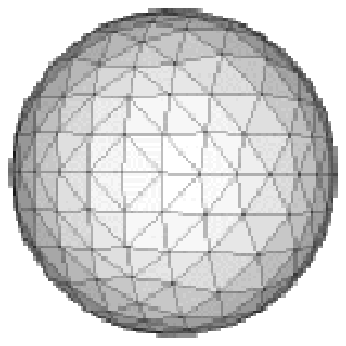}
    \hspace{1pc}
    \includegraphics[width=0.3\textwidth]{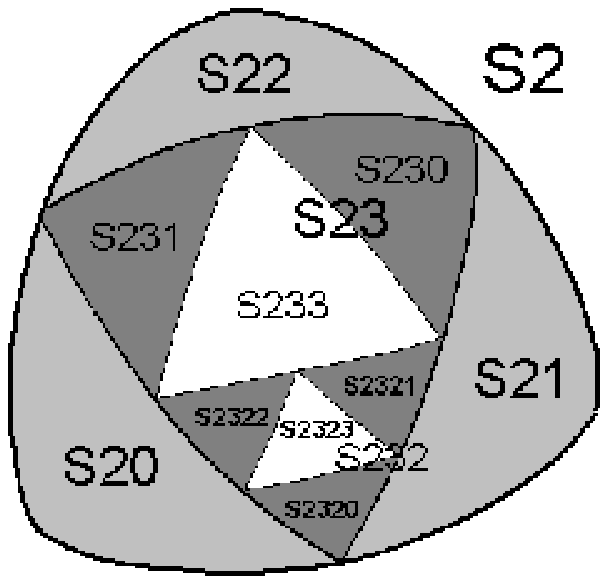}     
  \end{center}
  \caption{Recursive subdivision in HTM}
  \label{fig:mesh}
\end{figure}

HTM~\cite{KST01} maps triangular regions of the sphere to unique identifiers.
The technique for subdividing the sphere in spherical triangles is a recursive
process. At each level of the recursion, the area of the resulting triangles is
roughly the same (see figure \ref{fig:mesh}). In areas with a larger data density, the
recursion process can be applied with a greater level of detail than in areas
with lower density. The starting point is a spherical octahedron which
identifies 8 spherical triangles of equal size.

HEALPix~\cite{GHW98} is a curvilinear subdivision of the sphere in
quadrilaterals (pixel) of equal area (but variable shape). The contour of a
pixel is defined by the equation $cos\ \theta = a+b\times \phi$ on the equator
and by $cos\ \theta=a+b/ \phi^2$ on the polar regions. This structure makes more
efficient the execution of operations typically performed on the sky maps
including: convolution with local and global kernels, Fourier analysis with
spherical harmonics, nearest-neighbor searches.

%=============================================================
\subsection{KD-tree and its variants}
%=============================================================
The KD-tree~\cite{Bent75} is a binary tree that stores points of a
$k$-dimensional space. In each internal node, the KD-tree divides the
$k$-dimensional space into two parts with a $(k-1)$-dimensional hyperplane. The
direction of the hyperplane, that is the dimension on which the division in
performed, alternates between the $k$ possibilities from one tree level to the
following one. The subdivision process is recursive and terminates when the size
of a node (its longer side) or the number of points contained into it is below a
certain threshold. Given $N$ data points, the average cost of an insertion
operation is $O(\log_2 N)$. The tree structure and the resulting hierarchical
division of the space depends on the \emph{splitting rule}.

%% \begin{figure}
%% \begin{center}
%% \includegraphics[width=0.7\textwidth]{splits}
%% \end{center}
%% \caption{Splitting rule examples: standard split, midpoint split and sliding
%%   split}
%% \label{fig:splits}
%% \end{figure}

A drawback of KD-trees is that they have to be completely contained into the main
memory. With large datasets this is not feasible. KD-B-trees~\cite{Rob81} and
hB-trees~\cite{LS90} combine properties of KD-trees and B-trees to overcome
this problem.

%=============================================================
\subsection{R-tree and its variants}
%=============================================================
The R-trees~\cite{Gutt84} are hierarchical data structures meant to efficiently
index multidimensional objects with a spatial extent. They are used to store not
the real objects but their minimum bounding box (MBB). Each node of the R-tree
corresponds to a disk page. Similar to B-trees, the R-trees are balanced and they
guarantee an efficient memory usage. Due to the overlapping between the MBBs of
sibling nodes, in an R-tree a range query can require more than one search path
to be traversed.

To solve the overlapping problem, the R$^+$-tree access method introduced
in~\cite{SRF87} uses a clipping operation to avoid the intersection between
intervals at the same tree level. Objects intersecting more than one MBB at a
specific level are clipped and copied in several pages. This way, a single
search path is traversed for an exact match query. However, insertion operations
are more complex.

In the R-trees, search performances depend on the insertion algorithms.
In~\cite{BKSS90} an improved version of the R-tree, called R$^*$-tree, has been
proposed. This version uses a new insertion policy which significantly improves
search performances. The main target of this policy is to minimize the
overlapping between MBBs of sibling nodes to reduce the number of search path to
be traversed during a query operation.

%=============================================================
\section{Clustering algorithms on large datasets}
%=============================================================
Clustering algorithms have to locate regions of interest in which to perform more
detailed analysis and point out correlations between objects. An important
issue, in large datasets, is the efficiency and scalability of the clustering
algorithms with respect to the dataset size.

Many scalable algorithms have been proposed in the last ten years, including:
BIRCH~\cite{ZRL96}, CURE~\cite{GRS98}, CLIQUE~\cite{AGGR98}.

In particular, BIRCH is a hierarchical clustering algorithm. The main idea
behind the algorithm is to compress data into small subclusters and then to perform
a standard partitional clustering on the subclusters. Each subcluster is
represented by a \emph{clustering feature} which is a triplet summarizing
information about the group of data objects, that is the number of points
contained into the cluster and the linear sum and the square sum of the data
points. This algorithm has a linear cost with respect to the number of data
points.

CURE is an hierarchical agglomerative algorithm. Instead of using a single
centroid or object, it selects a fixed number of well-scattered objects to
represent each cluster. The distance between two clusters is defined as the
distance between the closest pair of representatives points and at each step of the
algorithm, the two closest clusters are merged. The algorithm terminates when
the desired number of clusters is obtained. To reduce the computational cost of
the algorithm, these steps are performed on a data sample (using suitable
sampling techniques). Its computational cost is not worse than the BIRCH one.

CLIQUE has been designed to locate clusters in subspaces of high dimensional
data. This is useful because generally, in high dimensional spaces, data are
scattered. CLIQUE partitions the space into a grid of disjoint rectangular units
of equal size. The algorithm is made up of three phases: first, it finds
subspaces containing clusters of dense units, than identifies the clusters,
and finally generates a minimum description for each cluster. Also this
algorithm scales linearly with the database size.

\section{Supervised learning and classification}
Classification algorithms are required in order to identify objects belonging to known
classes. In case of scientific (and in particular astrophysics) data, care has to
be taken on the interpretability of the classification results. For this reason,
one of the most popular methods to classify scientific data (in addition to neural
networks) is the algorithm based on decision trees~\cite{Quin86}. In fact, with
this method, the learning algorithm produces a binary tree which performs the
classification by means of value ranges on the data attributes.

Recently, Support Vector Machines (SVMs) are an active research domain within
the field of machine learning. 

\subsection{SVM for classification and novelty detection}
SVM and the related kernel methods are becoming popular for data mining
tasks like classification, regression and novelty detection. This approach is
systematic, reproducible, and properly grounded by statistical learning
theory~\cite{BC00}.

In its simplest form, an SVM is able to perform a binary classification finding
the ``best'' separating hyperplane between two linearly separable classes. There
are infinite hyperplanes properly separating the data. So, the SVM
finds this plane maximizing the distance, or \emph{margin}, between the
\emph{support} planes for each class (see \cite{Vap95} for theoretical
foundations). A plane supports a class if all points in that class are on one
side of that plane. This problem is formulated as a quadratic programming
problem (QP) and can be solved by effective robust algorithms. If the data
is not linearly separable, \emph{slack} variables are introduced into the QP
problem to accept outliers. Finally, a further non-linearity is introduced using
kernel functions (satisfying the Mercer's condition~\cite{Vap95}) to map data to
a higher dimensional space.

In many real problems, the task is not classifying but novelties or anomalies
detecting. In astrophysics, possible applications are the research of anomalous
events or new astronomical sources. An approach is modeling the \emph{support}
of a distribution (rather than estimating the density function of the data). A
method to solve this problem is represented by the Support Vector Clustering
(SVC) algorithm~\cite{BHSV01}, in which data are mapped to a higher dimensional
space by means of a Gaussian kernel function. In the new space, the algorithm
finds the minimum sphere enclosing the data. The Mapping of the sphere to the original
input space generates a set of contours enclosing the data and corresponding to
the support of the distribution.

\section{Conclusions}
In this work we have studied some data management and mining issues related to
astrophysical data, aiming at a complete data mining framework. In particular,
we have justified the need for a data warehousing approach to handle
astrophysical data and we have focused on multidimensional access methods to
efficiently index spatial and multidimensional data. A second issue concerns
clustering techniques on large datasets, and we have discussed about some
scalable algorithms with linear computational complexity. Finally, we have
focused on classification algorithms introducing an increasingly popular method
named Support Vector Machine, whose applications include the tasks of classification,
regression and novelty detection.

\end{document}